\documentclass[aps,pre,eqsecnum, twocolumn, 28pt]{revtex4}
\usepackage{amssymb,amsfonts,amsmath,bm}
\usepackage[dvipdf]{color}
\usepackage{graphicx}
\usepackage[cp1251]{inputenc} 
\usepackage[T2A]{fontenc}

\usepackage[english,russian]{babel}

\usepackage{hyperref}

\usepackage{xcolor}

\begin{document}



\title{Mean pairwise distances in Rouse polymer subject to fast loop extrusion}

\author{Ilya Nikitin$^{1,2}$, Nikolay Masnev$^{1}$ and Sergey Belan$^{1,2}$}
\email{sergb27@yandex.ru}
\affiliation{Landau Institute for Theoretical Physics, Russian Academy of Sciences, 1-A Akademika Semenova av., 142432 Chernogolovka, Russia}
\affiliation{National Research University Higher School of Economics, Faculty of Physics, Myasnitskaya 20, 101000 Moscow, Russia}

\begin{abstract} 
We consider a model of a Rouse polymer extended by the mechanism of active loop extrusion.
The model is based on a kinetic equation that is valid provided that the extrusion rate is high enough and the resulting loop ensemble is sufficiently sparse.
Within the one-loop approximation of diagrammatic calculations, a semi-analytical method for determining the mean square physical distance between a pair of chain beads as a function of the contour distance between them is developed.
The model is based on a kinetic equation that is valid provided that the extrusion rate is high enough and the resulting loop ensemble is sufficiently sparse.
Within the framework of the one-loop approximation of diagrammatic calculations, a semi-analytical method for determining the mean square of the physical distance between a pair of chain sections as a function of the contour distance between them is developed.
The mean square of the physical distance and its logarithmic derivative as functions of the contour separation are plotted for different values of the equilibrium degree.
The results are compared with the case of frozen disorder of sparse loops.
\end{abstract}

\maketitle

\vskip \baselineskip

%
%
%
%

%
%


\textit{Introduction.} 
The ideal polymer with an ensemble of random loops is a productive theoretical model that reproduces the peculiarities of contact statistics observed in the experimental data of the chromosome conformation method for the genome of higher eukaryotes \cite{Belan_2022,Polovnikov_2023,Starkov_2024}.
In addition, this model was used to extract analytical predictions about the statistics of pairwise intrachromosomal distances for the same cell types \cite{Belan_2024}.
The success of this model is explained by the fact that it takes into account the process of loop extrusion by molecular motors, a molecular mechanism that affects the spatial organization of chromatin at the submegabase scale of genetic distances \cite{Sanborn_2015, Fudenberg_2016, Fudenberg_2017, Mirny_2019,Dekker_2024}.
As was hypothesized long ago \cite{Riggs_1990, Kimura_1999, Nasmyth_2001}, and then confirmed experimentally \cite{Ganji_2018, Davidson_2019, Kim_2019, Golfier_2020, Ryu_2020}, SMC-protein complexes such as cohesin and condensin, when binding to DNA, are capable of creating dynamically growing chromatin loops in the presence of ATP.


To date, most of the theoretical results available in the literature obtained within the framework of the discussed model corresponds to the case of frozen loop disorder \cite{Belan_2022, Polovnikov_2023, Starkov_2024, Belan_2024}.
The assumption of statistical equilibrium of the loop ensemble is justified if the typical time required by the motor complex to extrude a loop of a given characteristic contour length is small compared to its relaxation time \citep{Polovnikov_2023}.
For a theoretical description of the statistics of chromosome conformation in a wider range of potentially possible experimental conditions, it is necessary to develop models with nonequilibrium loops.
To our knowledge, there are no analytical or semi-analytical results concerning the conformational statistics of ideal polymers with an ensemble of nonequilibrium loops yet. The most relevant works to the issue under discussion investigate the growth of individual loops \cite{Starkov_2021}, thus focusing on fairly short time and contour scales.
In this paper, we present an analysis of pairwise distance statistics for an ideal (Rouse) chain in the presence of a fast extrusion process.
Namely, we determine the mean square physical distance between a pair of chain points as a function of the contour distance between them.


\textit{Model formulation.}
Within the simplest phenomenological model of polymer stochastic dynamics, which is known as Rouse model, a polymer is represented as a chain of $N\gg1$ harmonically interacting beads subject to thermal noise. 
The coordinates of the intermediate beads  obey the Langevine equation in the form
\begin{equation}
\label{Rouse_eq}
\frac{d\vec r_n}{dt}=\gamma (\vec r_{n+1}+\vec r_{n-1}-2\vec r_n)+\frac{1}{\zeta}\vec \xi_n(t),
\end{equation}
while for the beads at the end of the linear chain we obtain
\begin{eqnarray}
\label{Rouse_eq1}
\dot{\vec {r}}_1=\gamma (\vec r_{2}-\vec r_1)+\frac{1}{\zeta}\vec \xi_1(t)
\end{eqnarray}
and 
\begin{eqnarray}
\label{Rouse_eq2}
\dot{\vec {r}}_N=\gamma (\vec r_{N-1}-\vec r_N)+\frac{1}{\zeta}\vec  \xi_N(t).
\end{eqnarray}
Here $\vec r_n$ is the position of the $n$-th bead,
$\vec \xi_n(t)$ is the Langevin force,  $\gamma=k/\zeta$ is the kinetic coefficient given by the ratio of the harmonic force stiffness $k$ over the viscous friction coefficient $\zeta$. 
The thermal forces are characterised by zero mean value $\langle \zeta_{n\alpha}(t)\rangle=0$ and the correlation function 
\begin{eqnarray}
\label{noise0}
\langle \xi_{n\alpha}(t_1)\xi_{m\beta}(t_2) \rangle = 2k_BT\zeta\delta_{\alpha\beta}\delta_{nm}\delta(t_1-t_2),
\end{eqnarray}
where $k_B$ denotes the Boltzmann constant, $T$ is the temperature of bath, 
$\delta_{nm}$ and $\delta_{\alpha\beta}$ are the  Kronecker symbols with Latin indices $n$ and $m$ labeling the beads, and Greek indices  $\alpha$ and $\beta$ labeling the Cartesian components of the vectors.

Note that harmonic forces in Eqs. (\ref{Rouse_eq})-(\ref{Rouse_eq2}) has entropic origin \cite{GKh_1994}. 
The excluded volume interation is neglected so that the chain is ideal. 
Also, Rouse model neglects hydrodynamic interaction in modelling polymer chain dynamics. 

Now let us increase the complexity of the model by introducing an extrusion mechanism.
The loop extruding units bind polymer  chain at random moments with uniform probability of binding per a bead a spend in the bound state exponentially distributed time intervals. 

In real conditions, the activity of the loop-extruding motor complex can be blocked by its collision with other motor units present on chromatin or by CTCF proteins that act as directional barriers to the extrusion process \cite{Fudenberg_2017,Spracklin_2023}.

In the absence of interaction with other motor complexes and barriers, the average contour length of the loop produced by an individual motor complex during its binding to the chain is $\lambda=v_0k_{\text{off}}^{-1}$.

On the other hand, the average contour distance between the binding sites of two adjacent complexes present on the chain in the statistically steady state can be expressed as $d=k_{\text{off}}/k_{\text{in}}$.
Note that both $\lambda$ and $d$ are dimensionless quantities defining the average number of beads.
Therefore, the dimensionless ratio 
\begin{eqnarray}
\varphi=\frac{\lambda}{d},
\end{eqnarray}
can be used to quantify the degree of sparseness of the loop ensemble.
In what follows, we will assume that the condition $\varphi\ll1$ is satisfied, which allows us to neglect the interaction between motor complexes.
In addition, our model will lack barriers that could prevent loop growth.

Given the assumptions made, we can use $\lambda$ as an estimate for the actual average loop size.
Then the probability density of a random loop contour length $l=0,2,4,\dots$ is given by





In the absence of interaction with other extruders and barriers, the mean contour length of loop produced by an individual motor complex would be $\lambda=v_0k_{\text{off}}^{-1}$.
On the other hands, the mean contour separation between binding sites of two neughboring motors present in the chain in the statistically-stationary regime can be expressed as $d=k_{\text{off}}/k_{\text{in}}$.
Note that both $\lambda$ and $d$ are the dimensionless quantities which quantify the average number of beads.
Therefore, the dimensionless ratio 
\begin{eqnarray}
\varphi=\frac{\lambda}{d},
\end{eqnarray}
can be used to quantify the sparseness degree of the loop ensemble.
In what follows we will assume that the condition $\varphi\ll1$ is satisfied, so that one can neglect  inter-motors collisions.
In addition, our model lacks barriers that could prevent the growth of loops.

Given these assumptions, we can use $\lambda$ as an estimate for the actual mean loop size.
Then, the probability density of the random loop length $l=0,2,4,\dots$ is given by 
\begin{eqnarray}
    \rho(l)=\frac{2}{\lambda+2}\left(\frac{\lambda}{\lambda+2}\right)^{\frac{l}{2}}.
\end{eqnarray}
The discrete quantity $l$ takes only even values which implies that motors extrude loops in a symmetric manner \citep{Banigan_2020}.

Next, let us introduce the time  $\tau_{\text{relax}}=\frac{\lambda^2}{\pi^2\gamma}$ required for relaxation to statistically equilibrium conformation of the loop of contour size  $\lambda$,  the typical duration  of an individual cohesin-polymer binding event $\tau_{\text{ext}}=k_{\text{off}}^{-1}$, and the average time interval $\tau_{\text{free}}=(\lambda k_{\text{in}})^{-1}$ between successive bindings of cohesins to a given chain segment of length $\lambda$.
Then the dimensionless parameter 
\begin{equation}
\sigma\equiv \frac{\tau_{\text{relax}}}{\tau_{\text{ext}}},
\end{equation}
measures how far from equilibrium a typical cohesin-mediated loop is, while the parameter
\begin{eqnarray}
  \epsilon\equiv \frac{\tau_{\text{relax}}}{\tau_{\text{free}}}=\frac{\lambda}{d}\sigma  
\end{eqnarray}
measures the same for the arbitrary chosen chain segment of length $\lambda$.

\textit{Kinetic equation}.
Now let us consider the joint probability density function ${\cal P}(\vec r_1,...,\vec r_N;t)$ of the set of beads coordinates $\vec r_1,...,\vec r_N$.
The closed-form kinetic equation for this object can be obtained in the limit of strongly non-equilibrium rare loops implying smallness of two dimensionless parameters, $\sigma\gg1$ and $\varphi\ll 1$.
Indeed, the condition $\sigma\gg1$ allows us to choose an intermediate time scale $\Delta t$ which is  large compared to $\tau_{\text{ext}}$ but small compared to $\tau_{\text{relax}}$.
Then, on the one hand, the event of cohesin binding and subsequent extrusion can be considered as instantaneous, and, on the other hand, the distribution function changes weakly over time $\Delta t$.
This change itself consists of the effects of beads thermal diffusion, beads displacement under the action of elastic forces, and the extrusion-mediated chain reconfiguration. 
In the fast extrusion and rare loop limits, i.e. $\sigma\gg1$ and $\varphi\ll 1$, the later effect can be modelled as collapse of the $l$ neigboring beads to the point of cohesin binding.
This allows us to write the following kinetic equation   
\begin{widetext}
\begin{eqnarray}
\label{kinetic_eq}
&&\partial_t{\cal P}(\vec r_1,...,\vec r_N;t)=\kappa \sum_{n=1}^{N} \Delta_{r_n}  {\cal P}(\vec r_1,...,\vec r_N;t)-\gamma \sum_{n=2}^{N-1} \vec \nabla_{r_n}[(\vec r_{n+1}-2\vec r_n+\vec r_{n-1}){\cal P}(\vec r_1,...,\vec r_N;t)]+\\
&&+\gamma \vec \nabla_{r_N}[(\vec r_{N-1}-\vec r_N){\cal P}(\vec r_1,...,\vec r_N;t)]+\gamma \vec \nabla_{r_1}[(\vec r_{2}-\vec r_1){\cal P}(\vec r_1,...,\vec r_N;t)]-\\ 
&&-k_{in} N{\cal P}(\vec r_1,...,\vec r_N;t)+k_{in} \sum_{n=1}^{N} \sum_{l=0,2,4,...}\rho(l)\prod_{i\in[-l/2,+l/2]\setminus\{0\}}\delta(\vec r_{n+i}-\vec r_n) \int {\cal P}(\vec r_1,...,\vec r_N;t) \prod_{j\in[-l/2,+l/2]\setminus\{0\}} d^3\vec r_{n+j}. 
\end{eqnarray}
\end{widetext}
which is valid at $t\gtrsim \Delta t$.
Here $\kappa=k_BT\zeta$ denotes the diffusion coefficient of a bead.

\begin{widetext}
\begin{figure}[t!]
\centering
\includegraphics[scale=0.56]{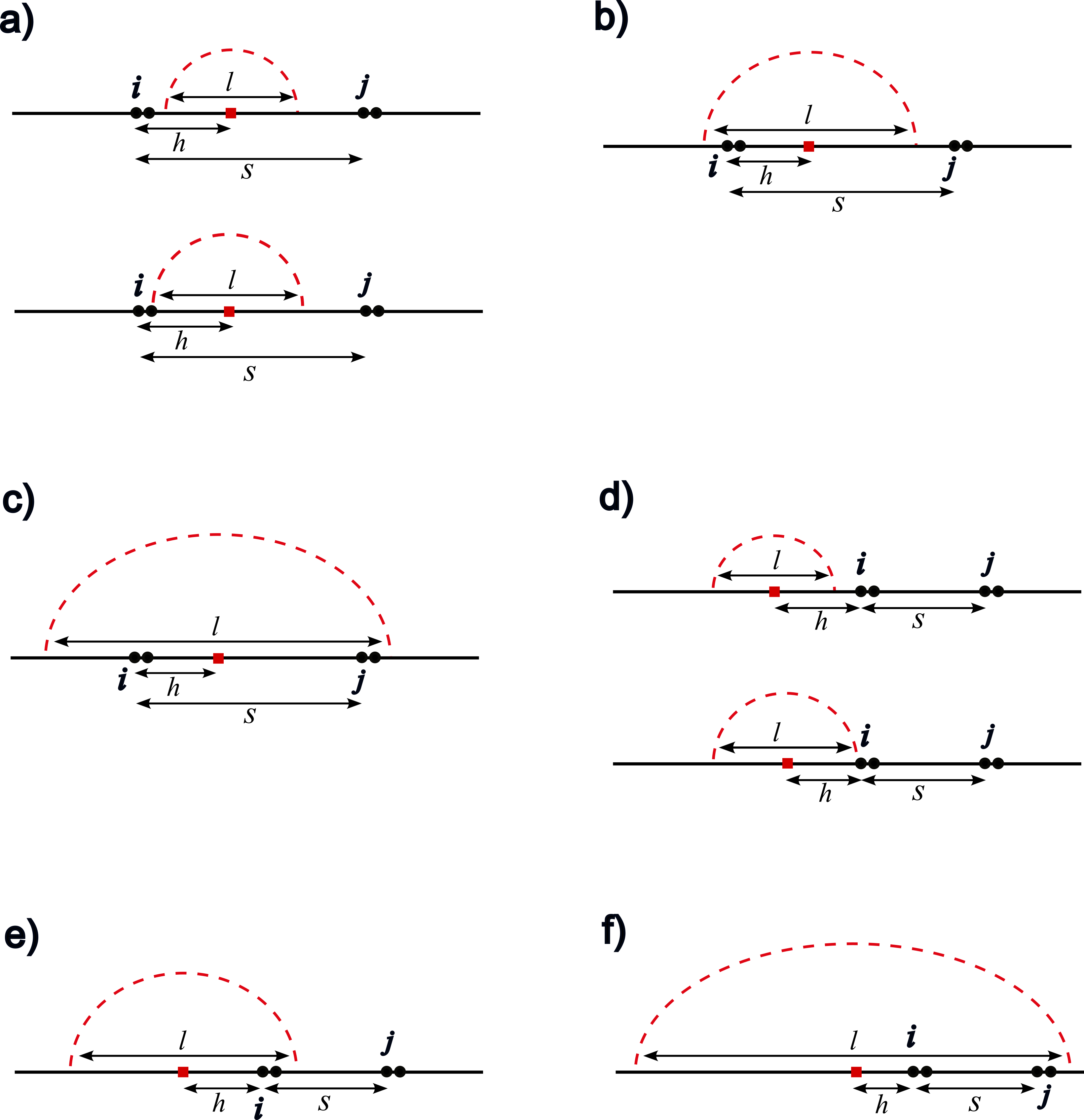}
\caption{Six types of diagrams contributing to the extrusion-induced terms in the set of equations  (\ref{set_of_eq_rouse})  on the pair correlation function $\mu(s)$.
Each diagram corresponds to a certain summand in the decomposition (\ref{represent1}).}
\label{fig:diagrams}
\end{figure} 
\end{widetext}

By setting $k_{in}$ to zero in Eq. (\ref{kinetic_eq}) we attain the Fokker-Plank equation for the probability density of beads coordinates in the standard Rouse model \cite{GKh_1994},
which can be derived from the set of Langevin equations (\ref{Rouse_eq})-(\ref{Rouse_eq2}). 
Next, the last but one term in the right-hand side of kinetic equation (\ref{kinetic_eq}) represents negative probability flux out of the each point of the chain phase space, while the last term represents the positive probability flux into the point associated with new chain configuration formed after fast extrusion activity of  motor unit. 
The double sum in the last term takes into account all possible realizations of the motor binding site and of the resulting loop length.  
It is easy to verify by integrating both sides of Eq. (\ref{kinetic_eq}) over beads coordinates that the normalization of the function ${\cal P}(\vec r_1,...,\vec r_N;t)$  is preserved in the course of time evolution.

\textit{Sketch of calculations.}
We aimed at determining the mean squared distance (MSD)
\begin{equation}
\langle \vec R^2(s)\rangle\equiv   \langle (\vec r_{j}-\vec r_i)^2 \rangle, 
\end{equation}
between a pair of beads $i$ and $j$ as a function of contour separation $s=j-i$ between them. 
In what follows we assume that  $1\ll i$, $i\le j$ and $N-j\gg 1$. 
Since $\vec r_{j}-\vec r_i=\sum_{k=i}^{j-1}\Delta \vec r_{k}$, where $\Delta \vec r_{k}$ is the vector connecting the neighboring beads $k$ and $k+1$, we can express this metric as 
\begin{eqnarray}
  &&\langle \vec R^2(s)\rangle = \sum_{k_1=i}^{j-1}\sum_{k_2=i}^{j-1}\langle \Delta \vec r_{k_1} \Delta \vec r_{k_2}\rangle=\\
  &&
  = b^2\sum_{k_1=i}^{j-1}\sum_{k_2=i}^{j-1} \mu(|k_2-k_1|)=\\
  && = s\mu(0) + 2\sum_{m=1}^{s-1}(s-m)\mu(m).
\end{eqnarray}
We introduced the notation 
\begin{eqnarray}
    \mu(|k_2-k_1|)=\frac{1}{b^2}\langle \Delta \vec r_{k_1} \Delta \vec r_{k_2}\rangle.
\end{eqnarray}
for the pair correlation function of the elementary separation vector 
and assumed statistical homogeneity of the chain along its contour. 
Here $b^2=3\kappa/\gamma$ - is the mean squared distance of the equilibrium Rouse chain in the absence of extrusion.

\begin{widetext}
\begin{figure}[t!]
\centering
\includegraphics[scale=0.5]{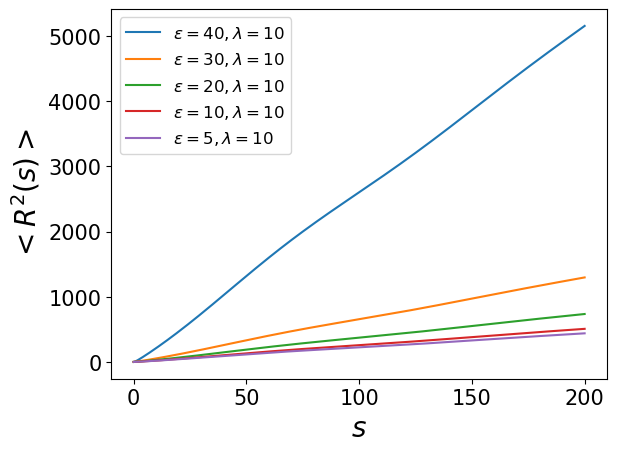}
\caption{Mean squared physical distance  $\langle R^2(s)\rangle$ in dependence of the contour separation  $s$ between pair of beads of the Rouse polymer chain in the presence fast loop extrusion  for different values  of loops non-equilibrium degree $\varepsilon$ and fixed mean loop size $\lambda$.}
\label{fig:diagrams}
\end{figure} 
\end{widetext}

To find $\mu(s)$ we multiply the kinetic equation (\ref{kinetic_eq}) by $\frac{1}{b^2\gamma}\Delta r_{i} \Delta r_{j}$ and integrate both sides over the coordinates of all beads.
Assuming statistically-stationary regime, $\partial_t{\cal P}=0$, we find the following  set of equations
\begin{widetext}
\begin{equation}
\label{set_of_eq_rouse}
\left\{\begin{array}{ll}
4-4\left(\mu (0) - \mu (1)\right)  - \frac{k_{in}}{\gamma} N \mu(0) + \frac{k_{in}}{\gamma} \sum\limits_{n=1}^{N} \sum\limits_{l=0,2,4,...}\rho(l)  G(0|l,n)=0, \\
\\
-2 +2(\mu(0)-2\mu(1)+\mu(2))- \frac{k_{in}}{\gamma} N \mu(1) + \frac{k_{in}}{\gamma} \sum\limits_{n=1}^{N} \sum\limits_{l=0,2,4,...}\rho(l)  G(1|l,n)=0,\\
\\
2(\mu(s-1)-2\mu(s)+\mu(s+1))- \frac{k_{in}}{\gamma} N \mu(s) + \frac{k_{in}}{\gamma} \sum\limits_{n=1}^{N} \sum\limits_{l=0,2,4,...}\rho(l)  G(s|l,n)=0,\ \ \ \text{for}\ \ \ s\ge 2, 
\end{array} \right.
\end{equation}
where
\begin{eqnarray}
\label{positive_contr}
&& G(s|l,n)= \frac{1}{b^2}\int  \prod_{k=1}^N d^3\vec r_{k} \Delta \vec r_i \Delta \vec r_j \prod_{m_1\in[-l/2,+l/2]\setminus\{0\}}\delta(\vec r_{n+m_1}-\vec r_n) \int {\cal P}(\vec r_1',...,\vec r_N';t) \prod_{m_2\in[-l/2,+l/2]\setminus\{0\}} d^3\vec r_{n+m_2}'.\ \ \ \ \  
\end{eqnarray}
\end{widetext}

In the extrusion-free case, $k_{in}=0$, we immediately find  $
\mu(s)=\delta_{s,0}$, which reflects statistical independence of  vectors $\Delta \vec r_1,\Delta \vec r_2,...,\Delta \vec r_{N-1}$ for equilibrium Rouse chain.

\begin{widetext}
\begin{figure}[t!]
\centering
\includegraphics[scale=0.45]{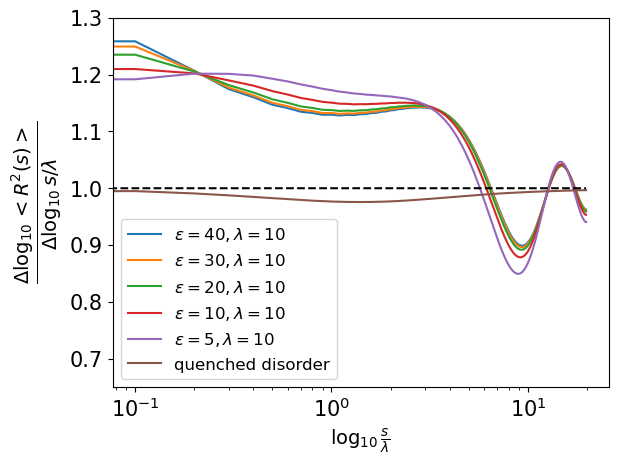}
\caption{Logarithmic derivative of the mean squared physical distance   $\langle R^2(s)\rangle$ in dependence of the contour separation $s$.
Different curves correspond to different values of the 
non-equilibrium degree of the loops $\varepsilon$.
In numerics we used the size $s_{\text{max}}=200$.}
\label{fig:logder}
\end{figure} 
\end{widetext}

\textit{Diagram representation}.
The last term in the left-hand sides of equations (\ref{set_of_eq_rouse}) can be represented as a sum of contributions 
\begin{eqnarray}
&&\frac{\alpha}{\gamma} \sum_{n=1}^{N} \sum_{l=0,2,4,\dots}\rho(l)  G(s|l,n)={\cal G}_a(s)+{\cal G}_b(s)+\ \ \ \ \ \\
\label{represent1}
&&+{\cal G}_c(s)+{\cal G}_d(s)+{\cal G}_e(s)+{\cal G}_f(s)={\cal G}_a(s)+{\cal G}_d(s),\ \ \ \ \ \ \ 
\end{eqnarray}
corresponding to different scenarios of mutual arragement of the beads $i$, $i+1$, $j$ and $j+1$ and the beads in the base of a motor-mediated loop.
These scenarios are presented in Fig.  \ref{fig:diagrams}  in the form of diagrams.
In appendix we show the contributions coming from the diagrams  (a) and (d) are linearly expressed via the correlation function $\mu(s)$.
Remaining diagrams ((b), (c), (e) and (f))  give zero contributions due to the structure of the expression (\ref{positive_contr}). 

Thus
\begin{eqnarray}
\label{represent2}
\frac{k_{in}}{\gamma} \sum_{n=1}^{N} \sum_{l=0,2,4,\dots}\rho(l)  G(s|l,n)={\cal G}_a(s)+{\cal G}_d(s).
\end{eqnarray}

\textit{Numerical solution}.
Let us pass to the matrix form of representation 
\begin{eqnarray}
\label{matrix_form}
    \hat A\cdot\vec \mu=\vec h,
\end{eqnarray}
The components of the vector $\vec \mu$ represent the values of the function $\mu(s)$ in discrete points $s=0,1,2,\dots$, i.e. $\vec \mu=(\mu(0),\mu(1),\mu(2),...,\mu(s_{\text{max}}))^T$,  $\vec h=(-4,2,0,...,0)^T$ - is the vector of free coefficients, and  $\hat{A}$ is the square matrix of dimension  $s_{max}+1\gg\lambda$ whose elements $a_{s,k}$  are given in additional material.
Indexes $s$ and $k$ enumerate rows and columns, respectively.
The large set of coupled equation (\ref{set_of_eq_rouse}) can be solved numerically.

 


\textit{Results.}
As can be seen from formulas (0.82), (0.84)-(0.98) in the additional materials, the amplitude of the extrusion-induced contributions to the coefficients of the system of equations (\ref{matrix_form}) is determined by the parameter
\begin{equation}
\chi=\frac{k_{in}}{\gamma}=\frac{\pi^2\varepsilon}{\lambda^3},
\end{equation}
and by the average loop length.
The Figures \ref{fig:diagrams} and \ref{fig:logder} present the numerical solution for the mean square of the physical distance $\langle R^2(s)\rangle$ and its logarithmic derivative $\frac{\log_{10}\langle R^2(s)\rangle}{d\log_{10}s/\lambda}$ for different values of the parameter $\varepsilon$ at fixed average loop size $\lambda$.

In the model of an ideal chain with frozen-in loop disorder, as shown in \cite{Belan_2024}, the mean squared physical distance as a function of the contour separation in the one-loop approximation is given by
\begin{equation}\label{msd}
\langle R^2(s)\rangle= l_{\text{eff}}s \left( 1 + \dfrac{\lambda}{d} f\left( \dfrac{s}{\lambda} \right) \right),
\end{equation}
where $f_{\text{MSD}}(z) = \frac{2}{3} \left(z^{-1}(1-e^{-z}) + {\cal E}_3(z)\right)-1$, and ${\cal E}_n(z)=\int_{1}^{+\infty}x^{-n}e^{-zx}dx$ is the exponential integral function.
The corresponding graph of the logarithmic derivative is given by one of the curves in the figure \ref{fig:logder}.

From the Fig. \ref{fig:logder} we see that the cases of quenched disorder and fast extrusion are qualitatively different. In the first case, the graph of the logarithmic derivative shows a single extremum: a minimum at the scale of the contour length of the order of $\lambda$.
In the second case, for not too small $\chi$, there is also a local minimum at this scale, but in addition to this, there is also a minimum at the scale of the order of $10\lambda$, followed by a maximum at a slightly larger contour distance.
This difference can potentially allow to estimate the degree of non-equilibrium of chromatin based on experimental data.


\textit{Conclusion.}
Based on the kinetic equation on the function of coordinates of the beads of the Rouse polymer chain in the presence of the mechanism of active extrusion of the loops, the average square of the physical distance as the function of the contour distance between the pair of beads was calculated.
A graph of the mean squared distance and of a logarithmic derivative of the mean squared distance was plotted.
The difference between the predictions of the present analysis and the previously known results obtained within the frozen loops model are discussed shown.

\acknowledgments

The work was supported by the Russian Science Foundation,
project no. 22-72-10052.
S.B. would like to thanks Leonid Mirny, Hugo Brandao and Anton Goloborodko for productive discussions on the subject on the present paper in 2017-2018.

{}

\end{document}